\documentstyle[prl,aps,epsf]{revtex}
\input epsf
\epsfverbosetrue
\begin{document}
\draft
\twocolumn[\hsize\textwidth\columnwidth\hsize\csname  
@twocolumnfalse\endcsname
\title{Superconductivity from a pseudogapped normal state: a 
mode coupling approach to precursor superconductivity}
\author{Jiri Maly, Boldizs\'ar Jank\'o, and K. Levin}
\address{The James Franck Institute, The University of Chicago, 5640
S. Ellis Avenue, Chicago IL 60637}
\date{\today}
\maketitle
\begin{abstract}
We derive a phase diagram for the pseudogap onset temperature 
$T^*$ (associated with the breakdown of the Fermi liquid state, due to 
strong pairing correlations) and the superconducting instability, 
$T_c$, 
as a function of variable pairing strength.  Our diagrammatic 
approach to the BCS - Bose-Einstein cross-over problem 
self consistently treats
the coupling between the single particle and pair 
propagators, and leads to  a continuous evolution of these 
propagators into the standard $T<T_c$ counterparts.  A rich structure 
is found in $T_c$ which reflects the  way in which the 
superconducting instability at $T_c$ is affected by the pseudogap 
$\Delta_{\rm pg}$.  An important consequence of Cooper-pair-
induced
pseudogaps is that  the magnitude of $T_c$ is sustained, even when 
$\Delta_{\rm pg}>T_c$. 
\end{abstract}
]
\makeatletter
\global\@specialpagefalse
\def\@oddhead{
REV\TeX{} 3.0\hfill \hfill Levin Group Preprint, 1997}
\let\@evenhead\@oddhead
\makeatother

It is generally agreed that the pseudogap state of the underdoped 
cuprates represents some type of pairing above $T_c$ which is 
postulated to derive from neutral spinon pairs \cite{RVB}, spin
\cite{Chubukov} or charge \cite{CDW}
density wave states, or from some form of  ($2e$) Cooper pairing 
which foreshadows the ultimate superconducting state. 
Each of these scenarios must not only address the nature of the
exotic (ie., pseudogapped) normal state but also the
transition from this state to its associated superconducting
instability.
Indeed, there is an extensive literature on the 
problem of establishing superconductivity in the presence of gaps in 
the electronic spectral function \cite{normal-gap}.
In this paper we investigate this 
issue  under the presumption that the normal state spectral function 
gap is associated with Cooper pairing above $T_c$.  Moreover, by 
self-consistently coupling the
single particle and pair properties we establish the nature and 
physics of the pseudogap state.  In this coupled scheme, we 
demonstrate how gapping 
in the electronic spectrum affects $T_c$ and argue that this $2e$ 
pairing is not particularly deleterious to the superconductivity, 
despite the relatively large size of the pseudogap
($\Delta_{\rm pg} > T_c $).  Our results are consolidated into  a phase 
diagram in which the pseudogap phase occupies a large temperature
range in the limit of moderately strong superconducting coupling.

An important contribution of this paper is to revisit the BCS - Bose 
Einstein cross-over problem \cite{NSR} using a conserving
diagrammatic 
``mode coupling'' formulation.
Above $T_c$ this approach leads to one and two electron Green's 
functions which are formally continuous with their below $T_c$ 
counterparts obtained in the standard theory of superconductivity. 
This approach also reproduces the expected limits associated with 
the extreme weak and strong coupling regimes of previous 
saddle point schemes \cite{Randeria}.  The intermediate coupling
regime is of particular importance to  the cuprates and Monte
Carlo simulations \cite{MonteCarlo} suggest that deviations
from Fermi liquid behavior are present there. Our previous diagrammatic
calculations \cite{Janko} indicate that long lived pairs
introduce a gap in the electronic spectral function by blocking 
available single particle
states around the Fermi surface. This gap correlates with  
a regime of $\it{resonant}$
pair scattering which forms a natural intermediate state between
the free fermions of the weak coupling, and the composite
bosons of the strong coupling limits. In the present fully self 
consistent theory, this resonant scattering will, itself, be modified by 
pseudogap effects.

We consider a generic Hamiltonian consisting 
of fermions in the presence of an attractive interaction, which may 
be modeled as  $V_{\bf k,k'} = 
g \varphi_{\bf k}\varphi_{\bf k'}$, where
$\varphi_{\bf k} = (1+k^2/k^2_0)^{-1/2}$ and $g<0$ is the coupling 
strength often expressed in units of 
$g_c=-4\pi/mk_0$ \cite{d-wave}. 
The BCS - Bose Einstein cross-over problem is characterized by 
coupled equations for $T_c$ and the chemical potential, $\mu$. The 
former derives from the Thouless criterion for the pair propagator or 
T-matrix, and the latter from the particle number equation, 
which involves the single particle propagator. In general, the 
full treatment of the coupling between the pair and single
particle propagators is referred to as ``mode coupling''.
In the present formulation the coupled equations are obtained 
following earlier literature on superconducting fluctuation 
effects \cite{PattonThesis}
\begin{eqnarray}
\Sigma^{ }_{{\bf k},i\omega_l} & = & T\sum_{{\bf q},\Omega_m}
t^{ }_{{\bf q},i\Omega_m}G^{(0)}_{{\bf q-k},i\Omega_m-i\omega_l}
\varphi^2_{{\bf k-q}/2} \label{Sigma} ,\\
t^{-1}_{{\bf q},i\Omega_m} & = & g^{-1} + T\sum_{{\bf k},\omega_l}
G^{ }_{{\bf k},i\omega_l}
G^{(0)}_{{\bf q-k},i\Omega_m-i\omega_l}
\varphi^2_{{\bf k-q}/2} \label{T-matrix-rep} .
\label{T-matrix}
\end{eqnarray}
where the Green's function is given by $G^{-1}_{{\bf k},i\omega_l} = 
G^{(0) -1}_{{\bf k},i\omega_l} - \Sigma^{ }_{{\bf k},i\omega_l}$,
$G^{(0) -1}_{{\bf k},i\omega_l} = i\omega_l-\epsilon_{\bf k}$,
$\Omega_m/\omega_l$ are the even/odd Matsubara frequencies, 
and
the electronic dispersion is $\epsilon_{\bf k}=k^2/2m-\mu$.

At high temperatures where mode coupling 
or ``feedback effects'' may be ignored \cite{Janko}, all Green's 
functions in Eqs.~\ref{Sigma} and \ref{T-matrix} may be
replaced by bare propagators. If we restrict our 
attention to moderately strong coupling constants, $g/g_c~\sim~1$
(associated with $ k_{\rm F}\xi/2\pi~\sim~1-10$), we  find  a 
Fermi liquid phase above a crossover temperature $T^*$.
The break-down of the Fermi liquid at $T_{\rm Res} \sim T^*$ 
coincides with the onset of resonant pair scattering. As the 
temperature is lowered below $T^*$ the gap in the spectral
function grows as the resonant states become
more long lived.  Finally, at $T_c$, coherent pairs condense into
a superconducting ground state.
The value of $T_c$ is obtained by solving
Eqs.~\ref{Sigma} and \ref{T-matrix}, which, at $T_c$, become
\begin{eqnarray}
\Sigma^{ }_{{\bf k},\omega} & = &
-\Delta_{\rm pg}^2\,\varphi^2_{\bf k}\,G^{(0)}_{{\bf k},-\omega} , 
\label{Sigma-Tc} \\
t^{-1}_{{\bf 0},0} & = &
g^{-1} + \sum_{\bf k}\frac{1-2f(E_{\bf k})}{2E_{\bf 
k}}\,\varphi^2_{\bf k}.
\label{Thouless-crit}
\end{eqnarray}
where  
$E_{\bf k} = \sqrt{\epsilon^2_{\bf k}+\Delta_{\rm pg}^2\varphi^2_{\bf k}}$.
The pseudogap parameter is, at arbitrary temperatures, defined to be
\begin{equation}
\Delta_{\rm pg}^2 = \sum_{\bf q}\int^{\infty}_{-\infty}
\frac{{\rm d}\Omega}{\pi}\, b(\Omega)
\,{\it Im}\, t_{{\bf q},\Omega} , \label{Delta-def}
\end{equation}
where $f(\omega),b(\omega)=(\exp(\omega/T)\pm1)^{-1}$.
It is evident that $\Delta_{\rm pg}^2$ in Eq.~\ref{Delta-def}
coincides with the square amplitude of pairing fluctuations,
$g^2<c^{\dagger}c^{\dagger}c^{ }c^{ }>$.

It should be noted that Eqs.~\ref{Sigma-Tc} and \ref{Thouless-crit}
provide an important validation for the diagrammatic scheme of 
Eqs.~\ref{Sigma} and \ref{T-matrix}.
These two equations show
that a pseudogap in the normal state 
spectrum preserves the overall ``BCS-like'' structure of the $T_c$ 
equation as well as the diagonal component of the single particle 
Green's functions. This continuity implies that, even though  these 
equations
are derived above $T_c$, they are formally 
identical to their one and two particle counterparts of the standard 
superconducting state \cite{Kadanoff}. This is in contrast to the use of 
fully renormalized Green's functions everywhere. 

In order to evaluate $T_c$ it is necessary to  further simplify 
Eq.~\ref{Delta-def}. As a first step, we investigate the T-matrix 
in the presence of a pseudogap. 
To capture the essential physics, we use a time dependent 
Ginzburg-Landau (TDGL) formulation 
\begin{equation}
gt^{-1}_{{\bf q},\Omega} = \tau^{ }_0 - (a'_0+ia''_0) \Omega + 
b^{ }_0\Omega^2
+ \xi^2_{\rm LG} q^2 ,
\label{LG-exp}
\end{equation}
where the TDGL parameters can be calculated by a formal expansion
of the full T-matrix at low frequencies and momenta.
For intermediate coupling 
and at sufficiently low $\Omega$ and
${\bf q}$, the T-matrix assumes the resonant form 
\begin{equation}
t^{ }_{{\bf q},\Omega} = 
\frac{ga^{-1}_0}{\Omega-\Omega_{\bf q}+i\Gamma_{\bf q}} ,
\label{T-resonant}
\end{equation}
where $a_0$ is predominantly  real. It follows
from Eq.~\ref{T-matrix},  that feedback effects, associated with a  gap 
in the single particle spectrum, lead to a gap in the 
imaginary part of the inverse T-matrix at $T_c$. This latter gap affects 
$\Gamma_{\bf q}$ in an important way:  when $\Omega_{\bf q}<\Delta_{\rm pg}$,
$\Gamma_{\bf q}\rightarrow0$.  This gap is slightly smeared out for 
temperatures between $T^*$ and $T_c$, but, nevertheless $\Gamma_{\bf q}$ 
remains small. Thus mode coupling leads to a stabilization of resonance 
effects, and, thereby an amplification of  pseudogap behavior. This TDGL 
formulation  is in contrast to results obtained from previous cross-over 
TDGL  theories
\cite{SadeMelo} which omit mode coupling, as well as from applications 
to  dirty superconductors \cite{PattonThesis}, where mode coupling is 
included, but gap effects are absent.

The above results  can now be applied to rewrite Eq.~\ref{Delta-def} 
and, thereby,  compute the superconducting instability
temperature, $T_c$. The value of $T_c$ is obtained 
from  a solution of the Thouless condition (Eq.~\ref{Thouless-crit})
and the number equation in the presence of 
the self-consistently determined pseudogap amplitude $\Delta_{\rm 
pg}$.  
The coupled equations for this transition temperature follow from 
Eqs.~\ref{Sigma-Tc}-\ref{Delta-def} and are given by
\begin{eqnarray}
1 + g\sum_{\bf k}\frac{1-2f(E_{\bf k})}{2E_{\bf k}}\,
\varphi^2_{\bf k}
& = & 0 , \label{Thouless-2}\\
2\sum_{\bf k}\left[v^2_{\bf k} + 
\frac{\epsilon_{\bf k}}{E_{\bf k}}\,f(E_{\bf k})\right] 
& = & n , \\
\sum_{\bf q}^{\Omega_{\bf q}<\Delta_{\rm pg}}
\frac{b(\Omega_{\bf q})}
{\left.\frac{\partial}{\partial\Omega}{\it Re}\,
t^{-1}_{{\bf q},\Omega}\right|_{\Omega_{\bf q}}}
& = & \Delta_{\rm pg}^2 ,\label{Psgap}
\end{eqnarray}
where $v^2_{\bf k}=(1-\epsilon_{\bf k}/E_{\bf k})/2$.

\begin{figure}
\epsfxsize=3in 
\hspace{0.5in}\epsfbox{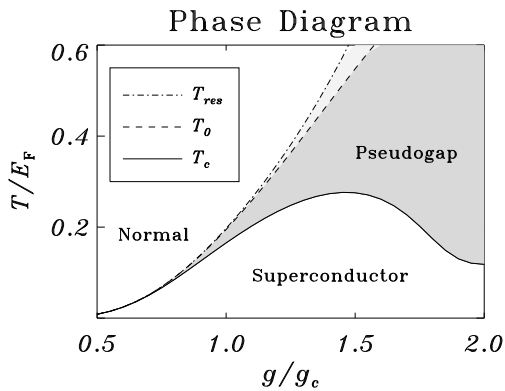}
\caption{Pseudogap phase diagram, which indicates the Fermi liquid breakdown 
co-rresponding to pair resonance onset, at $T_{\rm Res}$, the fully 
self-consistent $T_c$ and the transition temperature $T_0$ in the 
absence of mode coupling. }
\label{Phase}
\end{figure}

We solved Eqs.~\ref{Thouless-2}-\ref{Psgap} numerically
to obtain $T_c$ as a function
of $g/g_c$, along with the resonance onset temperature 
$T_{\rm Res}$, which we associate with $T^*$. 
Our results are plotted in Fig.~\ref{Phase}.
Also shown (dashed line) is the 
superconducting instability temperature $T_0$ computed 
in the absence of mode coupling. When there is an appreciable pseudogap,
$T_c$ and $T^*$ vary in an inverse fashion with $g/g_c$. This
is a consequence of the pseudogap which suppresses $T_c$, but is
not present at $T^*$. 
In this way 
the pseudogap regime is greatly enhanced by mode coupling contributions.

\begin{figure}
\epsfxsize=3in 
\hspace{0.5in}\epsfbox{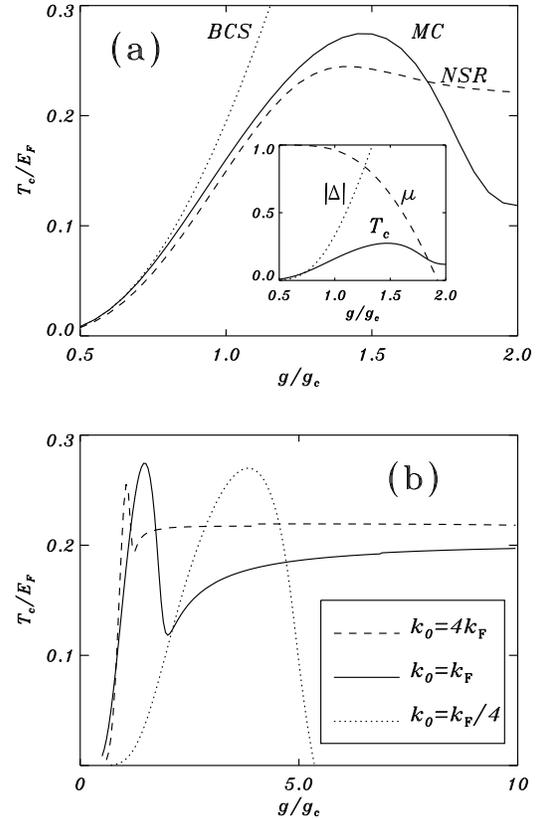}
\caption{(a) Comparison of the $T_c$ curves in BCS theory, the
Gaussian approximation of Nozi\`eres and Schmitt-Rink (NSR),
and our mode coupling theory (MC).  The inset plots the 
corresponding values of $\mu$, $\Delta_{\rm pg}$, at $T_c$.
(b) Variation of $T_c$ for three different pairing interaction ranges.} 
\label{Tc}
\end{figure}

The overall behavior of $T_c$ is compared with that obtained from 
the saddle point approximation, as well as that of strict BCS theory in 
Fig.~\ref{Tc}(a). Here the  inset plots the chemical potential 
and pseudogap function (at $T_c$), along with $T_c$.  It can be seen 
from the inset, that the maximum in $T_c$ is associated with $\mu 
\sim \Delta_{\rm pg}$ and the minimum with $\mu=0$.  Indeed, the 
complex  behavior of $T_c$ shown in Fig.~\ref{Tc}
can be understood on general physical 
grounds. A local maximum appears in the $T_c$ curve
as a consequence of a growing (with increased coupling)
pseudogap $\Delta_{\rm pg}$  in the fermionic spectrum which weakens 
the superconductivity. However, even as $\Delta_{\rm pg}$ grows,
superconductivity is sustained.  
In the present scenario, superconductivity is preserved by 
the conversion of an increasing fraction of fermions to bosonic 
states, which can then Bose condense. Once the fermionic conversion 
is complete ($\mu=0$), $T_c$ begins to increase again with 
coupling. This is a consequence of the decreasing pair size and 
concomitant reduction of the Pauli principle repulsion.

The behavior of $T_c$ on an expanded coupling constant scale, for 
different ranges of the interaction (parameterized by $k_0/k_{\rm 
F}$) is shown in Fig.~\ref{Tc}(b). The limiting value of
$T_c$ for large values of $g/g_c$ approaches the ideal
Bose-Einstein condensation temperature $T_{\rm BE}=0.218E_{\rm 
F}$
as $k_0\rightarrow\infty$ 
\cite{Haussmann}. 
The qualitative shape of the $T_c$ curve, however, is retained as 
long
as 
$k_0/k_{\rm F}$ is greater than about 0.5.
For 
larger range interactions, the solution disappears for some
$g/g_c$; then, when $\mu$ is sufficiently negative, the transition
reappears approaching a  continuously (with
${k_0/k_{\rm F}}\rightarrow0$) decreasing asymptote.

In order to implement the mode coupling scheme away from $T_c$, 
we introduce a two parameter fit to the full self-energy which is 
based on, and evolves smoothly into Eq.~\ref{Sigma-Tc} at $T_c$,
\begin{equation}
\Sigma^{\rm mc}_{{\bf k},\omega} = 
\frac{\Delta_{\rm pg}^2\varphi^2_{\bf k}}{\omega+\epsilon_{\bf k}
+i\gamma} .
\label{Sigma-ideal}
\end{equation}
This model makes it possible to do systematic numerics 
in the well established pseudogap phase, but it is not appropriate in 
a regime in which  the pseudogap breaks down as it evolves towards 
the Fermi liquid state. The instability of the Fermi liquid, or 
pseudogap onset $T^*$, is, however, addressed using the lowest order 
expansion of Eqs.~\ref{Sigma} and \ref{T-matrix}, as discussed above. 
The parameter  $\Delta_{\rm pg}$ in Eq.~\ref{Sigma-ideal} 
is defined from Eq.~\ref{Delta-def}, which is numerically 
implemented via
\begin{equation}
\Delta_{\rm pg}^2\,\varphi^2_{\bf k} = -\int^{\infty}_{-\infty}
\frac{{\rm d}\omega}{\pi}\, {\it Im}\, \Sigma^{ }_{{\bf k},\omega} .
\label{Del-def}
\end{equation}
Similarly,  the parameter $\gamma$ is obtained from the 
height of the 
peak in  $-{\it Im}\,\Sigma_{{\rm k},\omega}$ which is defined to be
$\Delta_{\rm pg}^2\varphi^2_{\bf k}/\gamma$. 
It should be stressed that 
at $T_c$ the parameter $\gamma$ in Eq.~\ref{Sigma-ideal}
becomes zero, as is implied by Eq.~\ref{Sigma-Tc}.
This results from the continuity 
of the pseudogap and superconducting phases, and can be seen 
explicitly by noting that $T_c$ occurs when the T-matrix becomes 
singular
at $\Omega=0$ and ${\bf q}=0$ (the Thouless criterion).
As a result, a divergent peak in ${\it Im}\,\Sigma_{{\bf k},\omega}$
occurs at $\omega=-\epsilon_{\bf k}$.

The coupled equations for $\Sigma_{{\bf k},\omega}$
and $t^{-1}_{{\bf q},\Omega}$, as a function of temperature, were 
solved 
numerically using the model self-energy of
Eq.~\ref{Sigma-ideal} and the TDGL approximation of Eq.~\ref{LG-exp}.
In a given iteration $\Delta_{\rm pg}$ and $\gamma$ were used to 
compute 
$t_{{\bf q},\Omega}$ via Eq.~\ref{T-matrix}
which was then approximated by Eq.~\ref{LG-exp}
for use in Eq.~\ref{Sigma}. The output of the latter was then used to 
extract $\Delta_{\rm pg}$ and $\gamma$ via 
Eq.~\ref{Sigma-ideal} and the procedure 
repeated until convergence. In this way $\Delta_{\rm pg}$, 
$\gamma$, 
$\Omega_{\bf q}$ , and 
$\Gamma_{\bf q}$ were all determined self-consistently as a 
function of temperature and coupling constant. 

\begin{figure}
\epsfxsize=3in 
\hspace{0.5in}\epsfbox{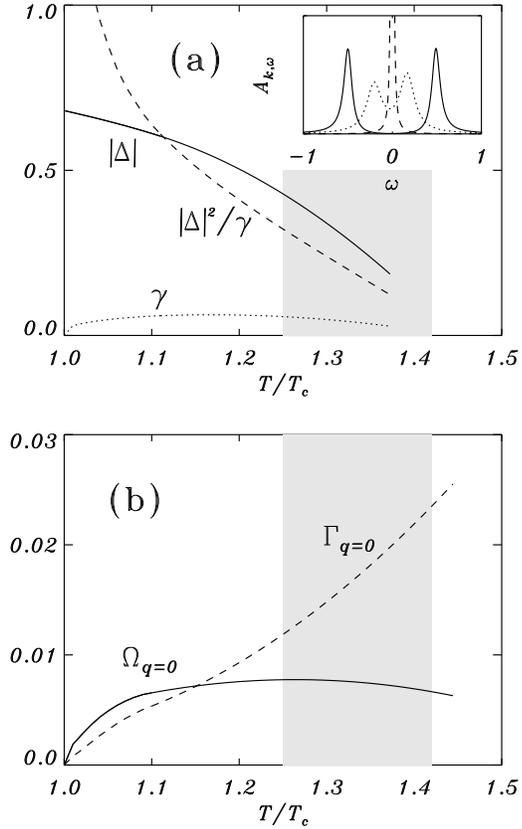}
\caption{Evolution of the parameters which characterize
the electronic  self-energy of Eq.~\ref{Sigma-ideal}, (a), and the 
pairing resonance (b). Shaded region represents the breakdown of 
Eq.~\ref{Sigma-ideal} as well as lowest order theory. The latter is 
valid to right of shaded region. The values of $\Delta_{\rm pg}^2/\gamma$ 
were divided
by a factor of 10 and  $g/g_c=1.2$. Spectral functions are plotted in the
inset, for the three temperature regimes.}
\label{Evol}
\end{figure}

To emphasize the important coupling between these electronic and 
pair properties, we compare the evolution of the single particle 
parameters
$\Delta_{\rm pg}$ and $\gamma$ with the pair parameters
$\Omega_{\bf q=0}$ and $\Gamma_{\bf q=0}$
in Fig.~\ref{Evol} for $g/g_c = 1.2$.
The associated spectral functions over this range of
temperatures are shown by the inset in Fig.~\ref{Evol}(a). The 
shaded 
regions indicate where the two parameter model self-energy of 
Eq.~\ref{Sigma-ideal} breaks down, as does the lowest order theory;
even though the 
pseudogap persists, the system begins to  cross-over towards the 
Fermi liquid. Thus the parameters $\Delta_{\rm pg}$ and $\gamma$ are 
indicated only for a limited range of temperatures.  The temperature 
for Fermi liquid onset, $T_{\rm Res}~\sim~T^*$, is just beyond the 
shaded region corresponding to the regime of validity of  the lowest 
order theory.  This temperature can be read off from the endpoint of 
the two curves in 
Fig.~\ref{Evol}(b).Moreover, the ratio $\Delta_{\rm pg}^2 / \gamma$ 
which represents the peak height of
${\it Im}\,\Sigma_{{\bf k},\omega}$,   appears to extrapolate 
smoothly 
from the mode coupling results to zero at 
$T_{\rm Res}$, at which point the Fermi liquid state sets in and
${\it Im}\, \Sigma_{{\bf k},\omega}$
changes character. As indicated in Fig.~\ref{Evol}(a),
the disappearance of the pseudogap arises via a reduction in 
$\Delta_{\rm pg}$, while the peak broadening remains relatively 
small.  This 
latter is a consequence of the extended lifetime of the pairs
due to a full ($s$-wave)  gap in the T-matrix.
This gap is also responsible for the relatively small 
values (when compared to conventional TDGL theory
\cite{PattonThesis})  of 
$\Gamma_{\bf q=0}$, which is associated with the inverse pair 
lifetime,
shown in Fig.~\ref{Evol}(b). At 
higher temperatures $\Gamma_{\bf q=0}$ and $\Omega_{\bf q=0}$ 
are 
plotted to join, after extrapolation, onto the results of the lowest 
order theory.

In summary, within a BCS - Bose Einstein cross-over picture, we have 
presented a quantitative phase diagram which compares the  
temperature onset of the pseudogap with the onset of a novel 
superconducting state, associated with pseudogapped fermions. 
Mode coupling effects, which were important for this analysis, 
considerably enhance the pseudogap regime. We 
have  demonstrated (for the $s$-wave case) that this pseudogap 
disappears with temperature, 
as it evolves into a Fermi liquid state, principally by a 
reduction in the gap size $\Delta_{\rm pg}$. While we have not 
established 
detailed connections to the cuprates, the inverse of 
$g/g_c$ in Fig.~\ref{Phase} can be loosely associated with hole 
concentration \cite{insulator}.
Of importance for the cuprates is the 
prediction that the effective inverse lifetime in the electronic
self-energy
$\gamma$, which can be deduced experimentally \cite{mike},  varies 
continuously to zero at $T_c$. Finally changes in our results 
associated with the $d$-wave, layered structure of the high $T_c$ 
systems can be anticipated: the nodal structure of $d$-wave pairs 
will 
weaken the gap in the T-matrix which played an important role in 
determining the detailed temperature evolution
of $\Delta_{\rm pg}(T)$ and $\gamma(T)$  
above  $T_c$. 
Moreover,   quasi-two 
dimensionality will considerably lower the energy scales and enhance 
the  pseudogap regime, particularly as the insulator is approached.  
Despite these omissions,  our physical picture of the 
interplay of the pseudogap and superconducting instability 
is 
expected to be qualitatively general, within a precursor superconductivity
scenario. 

We would like to thank A. Abanov, I. Kosztin, M. Norman and Y. Vilk 
for useful discussions.
This research was supported in part by the Natural Sciences and
Research Council of Canada (J.M.) and the Science and Technology
Center for Superconductivity funded by the National Science
Foundation under award No. DMR 91-20000.

\end{document}